\documentclass{solarphysics}
\usepackage[natbib,optionalrh]{spr-sola-addons}
\usepackage{url}            

\newcommand{\eg}{{\it e.g.}}

\usepackage{color}

\begin{document}
\begin{article}
  
\begin{opening}

\title{4$\pi$ models of CMEs and ICMEs}

%
\author{J.~\surname{Kleimann}
}

%
\runningauthor{J.~Kleimann}
\runningtitle{4$\pi$ models of CMEs and ICMEs}

%
\institute{Theoretische Physik IV, Ruhr-Universit\"at Bochum, \\
  44780 Bochum, Germany. email: \url{jk@tp4.rub.de}
}

\begin{abstract}
  Coronal mass ejections (CMEs), which dynamically connect the solar surface
  to the far reaches of interplanetary space, represent a major manifestation
  of solar activity. They are not only of principal interest but also
  play a pivotal role in the context of space weather predictions.
  The steady improvement of both numerical methods and computational resources
  during recent years has allowed for the creation of increasingly realistic
  models of interplanetary CMEs (ICMEs), which can now be compared to
  high-quality observational data from various space-bound missions.
  This review discusses existing models of CMEs, characterizing them by
  scientific aim and scope, CME initiation method, and physical effects
  included, thereby stressing the importance of fully 3-D ('$4\pi$') spatial
  coverage.
\end{abstract}

%

\end{opening}

%
\section{Introduction}

Coronal mass ejections (CMEs) are one of the most spectacular manifestations
of solar activity. Their mass and energy output, insignificant as they may
seem when compared to the Sun as a whole, are still able to severely distort
planetary magnetospheres even after they have been diluted considerably during
an expansion phase covering a distance of one AU or more, thus bringing the
solar influence to the outermost reaches of interplanetary space. \\
CMEs also relate to many other fields of solar and stellar physics. They are
intimately linked to flares
\citep[{\eg}][and references therein]{Shanmugaraju-etal-2011}, and often give
rise to strong shock fronts, which act as efficient accelerator engines for
solar energetic particles \citep[{\eg}][]{Reames-1999}.
It has also been recognized that CMEs play an important role for the
restructuring of the Sun's global magnetic field over the course of the solar
cycle \citep{Schwadron-etal-2008}. \\
Apart from these motivations to understand the CME phenomenon from a purely
scientific perspective, it has also become apparent in recent years that the
increasing complexity of modern technology, in particular communication
infrastructure, has also led to an increased vulnerability of this technology
to the adverse effects of {\em space weather}. The resulting
commercially-driven demand for high-precision forecasting tools has given
a continuing boost to research efforts in the field
\citep[][and references therein]{Baker-2002,Pulkkinen-2007}. Besides the
ensuing increased public interest in the subject, timely advances in computing
hardware and numerical algorithms have allowed for the development of
sophisticated large-scale models of CME dynamics, which are the subject of
this review. \\
The paper is structured as follows. After this introduction,
Section~\ref{s:model_classes} will discuss ways in which the vast spectrum of
existing CME models can be categorized, with special focus on their specific
strengths and possible weaknesses. Section~\ref{s:results} describes how
different types of models can be validated against obsevations, and
Section~\ref{s:conclusions} presents a concluding summary and a modest
suggestion for further action to improve on the comparability of results from
different models.

\subsection{What makes CME modeling a demanding task?}

To realistically model the evolution of a CME is a very demanding task for
several reasons:
\begin{enumerate}
\item The CME phenomenon spans vast temporal and spatial scales. Even
  disregarding the microphysics involved in, {\eg},\ flare reconnection
  (some $10^{-8}$~s, several meters) best described with kinetic approaches,
  a CME's life cycle from pre-eruption (when it extends over a small fraction
  of a solar radius $R_{\odot}$) to interplanetary expansion and finally
  merging with features like global interaction regions at the far reaches of
  the heliosphere extends over some six orders of magnitude in space and time,
  see \citet{Forbes-etal-2006} and in particular Figure~1 therein. For this
  reason, existing fluid-based models usually specialize on either
  initiation/eruption, interplanetary expansion, or interaction with
  corotating interaction regions, planetary magnetospheres, or other CMEs.
\item Even single CMEs show great variety in their morphology. Although during
  solar minimum, many of them originate from streamer blowouts and often
  exhibit the famous three-part "light bulb" structure consisting of a bright,
  semi-spherical front enclosing a dark cavity and a bright core 
  \citep{Illing-Hundhausen-1986}, the situation changes towards solar maximum;
  CME events then usually originate from active regions and tend to exhibit a
  much more irregular structure, which can differ markedly from the three-part
  textbook configuration. Even for prominence-related CMEs, the appearance of
  the three-part structure depends on the prominence location
  \citep{Cremades-Bothmer-2004}. The fraction of CMEs that cannot be
  classified into the subgroup of 'magnetic clouds' \citep{Burlaga-etal-1981}
  was estimated to be near 2/3 by \cite{Gosling-1990}, albeit with a marked
  variation throughout the solar cycle from almost none during some years of
  minimum up to $\sim${}60\% at maximum \citep{Li-etal-2011}. See
  \citet{Kilpua-etal-2011} for a recent review on magnetic cloud models and
  related multipoint observations. In their sample of almost 1,000 CMEs,
  \citet{Howard-etal-1985} were able to identify as many as ten morphological
  classes, and even this list is probably still far from exhaustive. On top of
  this, the large angular width of CMEs ($\approx 50^{\circ}$ on average, see
  \citet{St.Cyr-etal-2000}), combined with their high rate of occurrence
  especially during solar maximum, makes interaction among them likely, thus
  giving rise to an even wider spectrum of possible morphologies.
\item Attempts to reproduce observed events tend to be severely
  under-determined from observational side. To fully constrain a well-posed
  magnetohydrodynamic (MHD) model, initial and boundary conditions for
  density, pressure, and the velocity and magnetic field vectors must be
  specified; yet the initial (pre-eruptive) conditions are poorly known and
  usually rely on surface magnetograms and coronagraph images to constrain
  the magnetic field structure, possibly complemented by {\it in~situ}
  observations at individual locations.
\item CME propagation is an inherently three-dimensional (3-D) process. Even
  if the CME initially exhibits some form of spatial symmetry, its ensuing
  expansion in a complex magnetic environment will in any case break that
  symmetry, rendering modeling approaches with implied spatial symmetries
  problematic (see Section~\ref{s:symmetry}).
\end{enumerate}

\section{Model classifications}
\label{s:model_classes}

Both the complexity and the enormous range of scales covered by the CME
phenomenon make it necessary to devise separate models to address the
different stages of CME evolution, most notably the onset and eruption, the
ensuing phase of propagation and expansion, and a possibly following
interaction with other structures (planetary magnetospheres, other CMEs).

\subsection{Modeling CME onset...}

Since the kinetic energy of a CME is in most cases greater than what the
photosphere can provide during the timespan of the eruption, all current
models for the initiation phase assume a slow accumulation of energy, followed
by its rapid release due to loss of equilibrium. Existing analytical models
for this early phase are few in number and have to rely on simplifying
assumptions to keep the calculations manageable. \\
In the 2-D ideal MHD model of \citet{Forbes-Isenberg-1991}, the magnetic
energy of an initially stable filament is slowly increased by converging
advection of additional flux until a current sheet forms. Reconnection below
the filament can then lead to a 'catastrophe', {\it i.e.}, a sudden loss of
mechanical equilibrium which causes the filament to erupt.
\citet{Lin-etal-1998} extended this model and considered a finite curvature
along the invariant direction (thereby transforming the infinitely long
cylindrical filament into a force-free toroidal flux rope around the Sun).
While both models predict eruption as soon as the stored magnetic energy
exceeds a certain threshold, the now finite curvature force was shown to
yield a qualitatively different behavior, which now favors eruption for large
flux ropes. \\
Early numerical approaches to the simulation of solar eruptions relied on
re\-latively simple 2-D arcade geometries, in which a forced shearing motion
of the structure footpoints was employed to generate slow CMEs by driving the
system past a critical point in its parameter space
\citep[\eg][]{Mikic-etal-1988, Steinolfson-1991}. \\
The currently considered scenarios for CME outbreak largely fall into the
following groups:
\begin{enumerate}
\item The {\em mass loading} model relies on the notion that within a
  prominence, the equilibrium between magnetic tension and the gravity force
  of the mass above the structure allows for the slow accumulation of more
  mass, which merely causes the prominence to bend downwards under the
  increased weight of its load \citep{Fong-etal-2002, Zhang-Low-2004}.
  Eruption can occur when some restructuring of the magnetic field causes part
  of the material to suddenly drain away, causing a loss of equilibrium. These
  authors distinguish 'normal' and 'inverse' prominences based on their
  orientation with respect to the background field, and find that the normal
  variant is more likely to erupt. This finding has been confirmed numerically
  by \citet{Chane-etal-2006}, see Section~\ref{s:findinge}.
\item {\em Flux cancellation} models (also known as 'catastrophe models')
  start with a (usually twisted) arch-shaped flux rope which undergoes
  reconnective flux cancellation \citep{Martin-etal-1985} at its neutral line.
  This increases both the tube twist and its magnetic pressure at the expense
  of the overlying field, thus moving the equilibrium position to greater
  heights until eruption occurs because no further neighboring equilibrium
  exists. Notable numerical investigations of this process were carried out,
  in increasing order of complexity, by, \eg, \citet{Amari-etal-2000},
  \citet{Linker-etal-2003}, and \citet{Roussev-etal-2004}.
\item {\em Tether cutting} \citep{Moore-etal-2001} is very similar to flux
  cancellation, except maybe more impulsive and at slightly larger
  photospheric heights, cf.\ \citet{Chen-2011}. It starts with a single
  closed bipole (essentially a magnetic arcade) consisting of a twisted
  sigmoidal core and an 'envelope' of less twisted field lines. If the core
  axis is suitably aligned with respect to the neutral line, reconnection
  below the axis will weaken the tension of the envelope. This causes the core
  part to move upwards, thereby stretching the envelope field and leading to
  even more reconnection in the current sheet thus formed below until
  confinement becomes too weak to prevent the core from erupting. The model
  agrees well with x-ray images from the {\it Soft X-ray Telescope} (SXT) on
  board {\it Yohkoh}, but is deemed implausible on the basis of energy
  considerations \citep{Antiochos-etal-1999}. 
\item The {\em breakout model} \citep{Antiochos-etal-1999} starts with a
  quadrupolar arcade featuring an external X-type null point. The arcade is
  sheared and/or twisted by footpoint motions until the null is deformed into
  a current sheet at which reconnection sets in. This removes some of the
  overlying flux, which widens the current sheet further, thus leading to a
  runaway process which eventually causes the arcade to erupt. This model's
  crucial feature is that the ejected plasmoid is topologically detached from
  the arcade right from the beginning, which is a favorable condition to
  avoid the Aly-Sturrock constraint \citep{Aly-1984, Sturrock-1991}, see below.
  Corresponding MHD simulations have been carried out by
  \citet{MacNeice-etal-2004} and \citet{Lynch-etal-2004}.
\item Finally, two different variants of the {\em MHD kink instability} have
  been proposed as possible initiation mechanisms. Using the analytic field
  by \citet{Titov-Demoulin-1999} as initial configuration,
  \citet{Toeroek-Kliem-2005} showed that if this flux rope is twisted beyond
  a critical value, the twist is partially transformed into writhe, and a
  helical MHD kink instability sets in, as was already hinted at by an earlier
  stability analysis \citep{Hood-Priest-1981}. This may or may not lead to an
  eruption, depending on the strength of the overlying field
  \citep[\eg][]{Fan-2005, Rachmeler-etal-2009}. The kink instability thus not
  only addresses and explains the sudden release, but also the previous
  storage of magnetic energy, and can furthermore explain some related
  observations such as soft x-ray sigmoids \citep{Toeroek-etal-2004}. \\
  Second, \citet{Kliem-Toeroek-2006} analyzed the stability of a torus-shaped
  current loop against radial perturbations and the possibly ensuing torus
  instability, which they described as "a lateral kink instability distributed
  uniformly over the ring." They found that this instability may trigger the
  onset of a self-accelerating expansion along the major radius that can
  explain both the eruption of a flux rope (when treated as the upper half of
  a torus whose lower half is submerged below the photosphere) and the ensuing
  acceleration behavior of different types of CMEs, notably slow and fast
  ones.
\end{enumerate}
Common features of the last four models are the formation of a twisted flux
rope and the generation of a current sheet below the flux rope. What sets the
first and last model apart from the others is that they rely on purely ideal
MHD effects, and do not involve any reconnection (at least not as a trigger
element). Reconnection has been viewed as a convenient means to circumvent
the Aly-Sturrock constraint according to which any process which entirely
opens a force-free field to infinity would in fact increase the overall
magnetic energy and therefore could not drive an eruption. However,
\citet{Forbes-etal-2006} enumerate several other avenues to avoid the
constraint (such as fields being non-force-free, not simply connected, or
containing field lines disconnected from the Sun). Additionally,
\citet{Rachmeler-etal-2009} used a Lagrangian simulation scheme with no
(numerical or physical) resistivity whatsoever to demonstrate that
reconnection is not necessary to drive fast CME eruptions. This theoretical
finding is corroborated by the small but non-zero number of observed fast
CMEs showing no sign of flare association \citep{Marque-etal-2006}.
More details about the initiation and early propagation phase of CMEs,
especially of those originating from helmet streamers and prominences, can be
found in the reviews by \citet{Low-1996}, \citet{Forbes-2000}, and
\citet{Vrsnak-2008}, as well as in Chapter~8 of \citet{Howard-2011}. \\
Recently, \citet{Lin-etal-2010} analyzed the relevance of some of the
above-mentioned mechanisms for the kinematic evolution of two CME events and
found the breakout and catastrophe models to yield the best agreement with
their derived time profiles of height, velocity, and acceleration. This is
reminiscent to speculations by \citet{Howard-2011} suggesting that "...it is
likely that different types of CMEs are best described by different models.
Indeed, it is possible that most, if not all of the models [...] may be
appropriate to describe some CMEs under certain conditions."

\subsection{...and propagation}

\subsubsection{Trajectory mapping}

For the purpose of space weather forecasts, the three most crucial quantities
to be delivered by a model are the CME's trajectory, travel time, and
geo-effectiveness. To a first approximation, expansion is radial, implying
that only halo CMEs are likely to hit Earth. \citet{Schwenn-etal-2005} found
the respective rates of false and missing alarms for this correlation to be 15
and 20 percent, which underlines the need for more sophisticated approaches.
A first correction is the tendency of the Parker spiral magnetic field to
cause a slight westward deflection for fast CMEs, and a stronger eastward
deflection for the slower ones \citep{Wang-etal-2004}. From a sample of 841
CME observations using the {\it Large Angle and Spectrometric Coronagraph}
(LASCO) on board the {\it Solar and Heliospheric Observatory} (SOHO) mission,
\citet{St.Cyr-etal-2000} found 14\% of them to exhibit clear indication of
non-radial motion. \\
Observationally derived trajectories are often ambiguous because coronagraph
images suffer from projection effects, and the frequencies of the type~II
radio bursts associated with an eruption only provide the source heliocentric
distance (which furthermore depends on the assumed electron density profile).
Coronagraph data from the {\it Solar Terrestrial Relations Observatory}
(STEREO) mission \citep{Kaiser-2005} can be used for stereoscopic
reconstruction from image pairs \citep{Howard-Tappin-2008}. This, however, is
a well-defined problem only for curve-like objects \citep{Inhester-2006},
while for extended, diffuse objects like CMEs, some ambiguity remains. This
is usually resolved by manual identification of those points in image pairs
which are assumed to belong to the same point-like feature in physical space;
since this procedure uses two 2-D points to determine a singe 3-D point, it
is in fact an {\em over-defined} problem. For the small sample investigated
by \citet{Maloney-etal-2009}, these authors found their derived trajectories
to be consistent with quasi-radial expansion.

\subsubsection{Simple kinematic models}

Several authors have tried to deduce a CME's travel time $T$ in which it
covers a distance $R_{\rm s/c}$ as a function of its initial velocity $v_0$ by
fitting formulas like
\begin{equation}
  v_0 T + a T^2 /2 = R_{\rm s/c}
\end{equation}
to $(v_0,T)$ data pairs from actual events, thus using the required
acceleration $a$ (due to thermal pressure, magnetic forces, gravity, momentum
conservation \citep["snow plough effect",][]{Tappin-2006}, and aerodynamic
drag) as a free parameter. \citet{Gopalswamy-etal-2001} found a minimum
variation of
\mbox{$\Delta T \approx$ 10 h} for their sample if $a=0$
beyond 0.75~AU. The tentative inclusion of a drag term
\mbox{$a(v) \propto (v - v_{\rm sw})^2$} \citep{Cargill-2004} was shown to have
little influence on the method performance, though it is indeed observed that
for large distances, the CME speed will approach the speed $v_{\rm sw}$ of the
ambient medium \citep{Lindsay-etal-1999, Maloney-etal-2009}. Relating these
empirical $T=T(v_0)$ formulas to observed events yields large discrepancies,
which can be attributed to the inherent oversimplification of this method.
Therefore, it has to be concluded that both the CMEs themselves and the
interplanetary medium which they encounter are much too variable and
structured for simple fitting laws of this kind to yield more than rough
estimates. \\
As an intermediate step between one-parameter fitting and fully
self-consistent MHD simulations, spatially resolved kinematic models such as
the HAF model \citep{Hakamada-Akasofu-1982, Fry-1985} have been used to
predict the arrival times of shocks and the spatial structure of the inner
heliosphere as it is modulated by corotating interaction regions. This model
projects a (possibly time-dependent) boundary condition at
\mbox{$2.5 R_{\odot}$} outwards along stream lines, and has been calibrated
using direct (albeit only 1-D) MHD simulations \citep{Sun-etal-1985}.

\subsubsection{Numerical MHD propagation models}
\label{s:symmetry}

The numerical study of propagating CMEs is a challenging task mainly because
the need to track structures with details smaller than a solar radius across
at least 1~AU requires very high spatial resolution. On top of that, the
configuration has no apparent symmetry properties that could be exploited to
reduce computational expenses. During solar maximum, the Sun's global magnetic
field is inherently 3-D, and even during solar minimum, CMEs are never
observed to travel along the Sun's polar axis, thus breaking the rotational
symmetry of the field. \\
In the past, two strategies have been used to deal with this difficulty.
The first is to ignore the misalignment of expansion direction and magnetic
symmetry axis and to assume cylindrical symmetry anyway
\citep[\eg][]{Chane-etal-2006}. This implies a closed, torus-shaped CME
geometry which is not anchored on the Sun, a configuration that could
potentially be relevant for tube-shaped magnetic clouds. In their comparison
of 2-D versus 3-D models, \citet{Jacobs-etal-2007} conclude that propagation
models with cylindrical symmetry, however inadequate from a principal point of
view, can still provide useful and computationally inexpensive estimates,
which can then be used to set up a refined and symmetry-free follow-up
simulation. The general usefulness of such approaches requires that the
parameters are properly transferred between both geometries, which necessarily
requires several ambiguities to be resolved. \\
The second avenue is to use a fully 3-D model with sufficient resolution in
all three coordinate directions. The resulting vast increase in computational
expense can be moderated by the use of specially tailored grids, in particular
spherical grid geometries with a radial mesh spacing \mbox{$\Delta r$} which
increases with heliocentric radius $r$ \citep[\eg][]{Pomoell-etal-2011}. \\
For a global fluid simulation of CMEs, the numerical grid must be chosen
carefully to optimize the trade-off between the advantages and shortcomings of
different grid geometries. While spherical, Sun-centered grids are obviously
best adapted to the Sun's shape and thus greatly simplify the specification of
boundary conditions and the description of predominantly radial expansion, the
large variation in cell sizes may lead to undesirably low time steps, and the
necessary inclusion of the polar axis requires a delicate treatment of
coordinate singularities. The latter can be avoided by the use of so-called
overset grids, such as the Cubed Sphere \citep{Ronchi-etal-1996} or the
Ying-Yang grid \citep{Kageyama-Sato-2004}. These, however, require some form
of interpolation scheme to establish a seamless connection between the
sub-grids, which usually destroys the conservative properties of the code
within the interpolation region. For applications of these grids to 3-D solar
wind modeling see \citet{Feng-etal-2010,Feng-etal-2011}.  Cartesian
coordinates, which do not present such difficulties, remain popular among CME
modelers for exactly this reason, although the implementation of spherical
boundaries like the photosphere then either requires some form of weighted
interpolation procedure \citep[\eg][]{Kleimann-etal-2009} or a much increased
spatial resolution. The latter requirement becomes much less severe by the use
of adaptive mesh refinement techniques
\citep[\eg\ {\sc bats-r-us},][]{deZeeuw-etal-2000}. Logically rectangular
grids \citep{Calhoun-etal-2008}, which permit the smooth inclusion of curved
boundaries into an otherwise Cartesian grid geometry, harbor the potential to
combine the advantages of Cartesian and spherical coordinate grids, but have
until now not been used in any large-scale 3-D simulation with relevance to
space physics. \\
Another interesting approach is the use of multi-scale models
\citep[\eg][]{Riley-etal-2006}, which consist of a radially nested sequence
of grids, such that the outer boundary condition of a given grid is fed into
the next, coarser grid as an inner boundary condition of the latter. The fact
that the solar wind plasma flow becomes super-Alfv\'enic after only a few
$R_{\odot}$ frees these authors from the requirement to run the model
simultaneously on all sub-grids. The multi-scale approach is also a key
feature of the {\em Space Weather Modeling Framework}
\citep[SWMF, see][]{Toth-etal-2005}, which condenses a total of nine model
components (for the magnetosphere, inner heliosphere, radiation belts, etc.)
into a modular, versatile space physics-based simulation environment.

\subsection{Classification by aim and scope}

Existing models for the expansion/interaction phase of CMEs largely fall into
two categories:
\begin{enumerate}
\item 'principal' studies, which intend to investigate the relevance of a
  (usually small) set of free parameters on the resulting dynamics, and
\item 'realistic' simulations, which are aimed at a usable prediction of the
  development of actual events, and therefore need to include as much physics
  and data as possible.
\end{enumerate}
Consequently, these two types of models differ noticeably in the way in which
they implement boundary and initial conditions. On the one hand, principal
models tend to use a background solar wind which is either uniform
\citep[\eg][]{Vandas-etal-2002, Dalakishvili-K-etal-2011} or moderately
structured \citep[\eg][]{Odstrcil-Pizzo-1999, Manchester-etal-2004}, combined
with a dipolar, or at most quadrupolar magnetic field, and possibly amended
with a planar current sheet.
Realistic global-scale models, on the other hand, routinely rely on synoptic
magnetograms to specify a potential coronal magnetic field as a photospheric
boundary condition, from which a reasonable initial condition can be derived
\citep[\eg][]{Hayashi-etal-2006, Lugaz-etal-2007}. To fully constrain this
field solution, additional assumptions about the type of {\bf B} field need to
be made, the most popular being that the field is either potential
\mbox{($\nabla \times {\bf B} = {\bf 0}$)}, linearly force-free
\mbox{($\nabla \times {\bf B} = \alpha {\bf B}$)}, or non-linearly force-free
\mbox{($\nabla \times {\bf B} = \alpha({\bf r}) \ {\bf B}$)}.
For a review of strategies and algorithms to obtain force-free solutions see
\citet{Metcalf-etal-2008}; a critical assessment of their performance on
actual data has been presented by \citet{DeRosa-etal-2009}. \\
Alternatively, vector magnetograms have been used to calculate localized
non-linear force-free fields on the scale of active regions, from which
CME-like eruptions can be launched in a simulation \citep{Kataoka-etal-2009}.
Once the magnetic field is known, empirical formulas can be used to determine
suitable initial conditions for the remaining fluid quantities density,
velocity, and temperature \citep[\eg][]{Detman-etal-2006}.

\subsection{Method of initiation}

Given that the physical process which triggers an eruption is still not
conclusively settled, there is some freedom in the choice of methods to
numerically start this process. Density-driven models
\citep{Groth-etal-2000, Chane-etal-2006, Kleimann-etal-2009}
increase the pressure and/or density (and occasionally also the radial
momentum, thereby imposing an additional mass inflow
\citep[\eg][]{Keppens-Goedbloed-2000}) below a closed field line configuration
until the inward tension of the field can no longer confine the growing
internal pressure and gives way to a violent expulsion of the accumulated
plasma blob. Similarly, additional magnetic flux can be artificially pushed
through the photosphere \citep[\eg][]{Fan-Gibson-2007, Fan-2011}. Note that
this latter approach must be carefully distinguished from flux emergence
models, which are used to study the physical process of buoyant flux tubes
emerging from the convection zone. The initial field may be further
destabilized by, \eg, local magnetic reconnection
\citep[\eg][]{Forbes-Priest-1995, Chen-Shibata-2000}, or by an initial force
imbalance, \eg { } due to buoyancy. Popular initial setups of the latter kind
include the self-similar expanding flux rope by \citet{Gibson-Low-1998}, which
has been employed by, \eg, \citet{Manchester-etal-2004} to initialize
simulations of CMEs, just like the famous Titov-D\'emoulin flux rope
\citep{Titov-Demoulin-1999}. \\
Finally, the required instability can also be produced by shearing the
photospheric footpoints of an arcade until it becomes unstable and is forced
to open up spacewards. The effect of shearing motions on solar magnetic
structures has been investigated theoretically \citep{Low-1977},
observationally \citep{Deng-etal-2001}, and numerically
\citep[\eg][]{Mikic-Linker-1994, Jacobs-etal-2006}. \\
From these methods, the density-driven one is special in that it is not
supposed to mimic a process that is actually assumed to happen on the Sun;
moreover, its justification lies in the pragmatic notion that if the focus is
on the propagation phase, the details of initiation are of secondary
importance, thus justifying the use of a method that is easy to implement and
produces a relatively realistic outcome. The same rationale was also given by
\citet{Pomoell-etal-2008}, who start their Cartesian 2-D simulation using a
set of line currents similar to the one employed by \citet{Chen-etal-2002},
and then apply an artificial volume force to the interior of the detached flux
rope to pull the core region upwards (which requires the trajectory of
interior fluid elements to be dynamically traced during the acceleration
phase).

\subsection{Treatment of the energy budget}
\label{s:simple_heating}

Despite almost six decades of undiminished efforts to understand the heating
of the solar corona, this vital issue is still far from being settled
\citep{Cranmer-2002,Marsch-2006}. For practical reasons, models seeking to
describe a CME's propagation in the corona and inner heliosphere have to adopt
a formalism to model the distribution and release of energy, which opens up
another possibility to classify them. \\
The simplest one would be to link the plasma pressure $p$ and mass density
$\rho$ via an adiabatic closure relation
\begin{equation}
  \label{eq:ad-close}
  p \sim \rho^{\gamma} \ .
\end{equation}
In this case, again the simplest (albeit also the most unrealistic) choice
would be to assume an (almost) isothermal plasma by setting the adiabatic
index $\gamma$ to unity, or a value slightly larger than unity, such as 1.05.
Using the physically appropriate value of $5/3$ in conjunction with
Equation~(\ref{eq:ad-close}) is usually not feasible because the effect of
adiabatic cooling would quickly reduce the temperature down to unrealistically
low values. This problem is sometimes mediated by using a spatially varying
adiabatic exponent, either as an explicitly prescribed dependence
$\gamma=\gamma({\bf r})$ \citep[\eg][]{Fahr-etal-1977, Wu-etal-1999}, or
derived from energy considerations based on the Bernoulli equation
\citep{Cohen-etal-2007} or from interpreting the low value of $\gamma$ close
to the Sun in terms of internal energy being stored as additional degrees of
freedom in large-scale turbulence \citep{Roussev-etal-2003}. The obvious
downside of using unphysically low adiabatic indices is the resulting
inability of the model to correctly reproduce the properties of shocks, which
may or may not be acceptable in a given situation, but which must in any case
be kept in mind when interpreting the results thus obtained. \\
If $\gamma=5/3$ holds throughout the considered volume of space, a full energy
equation for the total energy density $e$ like
\begin{equation}
  \partial_t e + \nabla \cdot \left[ \left( e+p+ \|{\bf B}\|^2/2 \right)
    {\bf u} -  \left( {\bf u} \cdot {\bf B} \right) {\bf B} \right]
  = ({\bf g} \cdot {\bf u} +Q)\rho
\end{equation}
(in which ${\bf u}$ denotes the flow velocity and  ${\bf g}$ the acceleration
due to gravity) with heating source term $Q$ has to be used. Popular choices
for $Q$ include {\it ad~hoc} heating functions like
\begin{equation}
  Q({\bf r}) = q({\bf r}) \left[T_0-T \right]
\end{equation}
\citep{Manchester-etal-2004}, where $T_0$ is a spatially dependent "target
temperature", which is fine-tuned to reproduce a realistic quiet-Sun
temperature distribution. The main shortcoming of this approach is that it is
obviously biased towards the target temperature, or in other words, it will
systematically underestimate the temperature change which is brought about by,
\eg, the passage of a CME. Another, more recent approach is that of
\citet{Pomoell-etal-2011}, who employ an energy equation
\begin{equation}
  \partial_t (p / \rho^{\gamma})
  + {\bf u} \cdot \nabla \left( p / \rho^{\gamma} \right) = S
\end{equation}
and first run a quiet-Sun simulation using $\gamma=1.05$ and $S=0$ (and no
CME), followed by the 'main' run using $\gamma=5/3$ and
\begin{equation}
  S = {\bf u}_1 \cdot \nabla \left( p_1 \left/ \rho_1^{5/3}\right. \right)
\end{equation}
where the subscript 1 denotes values from the $\gamma=1.05$ run. This
procedure guarantees that both runs share the same stationary (quiet-Sun)
state.

\subsection{Alfv\'enic wave heating}

As an alternative to the pragmatic heating functions of
Section~\ref{s:simple_heating}, attempts have been made to amend the
underlying solar wind model with a more physical heating process. A promising
candidate for such a mechanism is heating by Alfv\'en waves, which are
generated at photospheric levels, then travel outward along magnetic field
lines while being shifted towards an upper limit frequency $f_{\rm h}$ at which
they dissipate and deposit their energy as heat. In its full form, this scheme
requires an entire wave power spectrum $P(f,{\bf r},t)$ to be modeled for each
position ${\bf r}$. A dynamic equation for the temporal change of $P$ has to
be solved for each frequency $f$, and the waves then couple back to the MHD
system via:
\begin{enumerate}
\item the accelerating negative gradient of the wave pressure
  \begin{equation}
    p_{\rm w} \sim \int_{f_0}^{f_{\rm h}} P(f) \ {\rm d}f
  \end{equation}
  representing the magnetic pressure of the fluctuations, and
\item a wave heating term
  \begin{equation}
    Q_{\rm w} = F(f_{\rm h},{\bf r}) - P(f_{\rm h}({\bf r}),{\bf r})
    \left[ {\bf u} \pm {\bf v}_{\rm A} \right]
    \cdot \nabla f_{\rm h}({\bf r})
  \end{equation}
  in which the so-called cascading function $F=F(P,f)$ governs the spectral
  evolution of $P$, and ${\bf v}_{\rm A}$ is the local Alfv\'en velocity.
\end{enumerate}
More details can be found in the review by \citet{Fichtner-etal-2008}. This
formalism has been successfully applied to purely radial models of the
quiet-Sun solar wind \citep{Laitinen-etal-2003}, and appears to be a promising
alternative to the phenomenological heating functions discussed in
Section~\ref{s:simple_heating}, although the extension to the generally more
complicated magnetic field structures encountered in symmetry-free CME
simulations is clearly non-trivial. First results obtained with a wave-heated
solar wind model (albeit relying on the scalar wave energy densities
$\varepsilon_{\pm}$ parallel ($+$) and anti-parallel ($-$) to ${\bf B}$ rather
than the full spectral information) have been presented by
\citet{vanderHolst-etal-2010}.

\section{Connecting to observations}
\label{s:results}

As can be expected from their very different aims, 'principal' and 'realistic'
models give qualitatively different types of results; one could also say that
they provide answers to different types of questions.

\subsection{Selected findings from 'principal' models}
\label{s:findinge}

Since several analytical CME models rely on the simplifying assumption of
self-similarity \citep{Farrugia-etal-1995, Gibson-Low-1998,
  Nakwacki-etal-2008, Wang-etal-2009}, it is a vital question whether this
property can be confirmed numerically. \citet{Dalakishvili-R-etal-2011}
considered an idealized model for a cylindrical magnetic cloud, and
numerically confirmed that an initially self-similar configuration maintains
this property at later stages. Using a slightly more realistic 3-D CME model,
\citet{Kleimann-etal-2009} compared the derived time profiles of density and
magnetic field strength at different heliospheric distances, and also found
indication of self-similar evolution at least in the early phase of
propagation, which in this case covered a mere 10 $R_{\odot}$. This is
consistent with observations which show a CME's cone angle to be approximately
constant in time \citep{Schwenn-etal-2005}, although other simulations have
been performed in which the evolution starts to depart from self-similarity
after several tens of $R_{\odot}$
\citep[\eg][]{Manchester-etal-2004_ip, Chane-etal-2006}. As was demonstrated
by \citet{Riley-Crooker-2004}, these findings can be reconciled by observing
that the passive advection of a structure in a spherically diverging flow
causes the aspect ratio (azimuthal over radial extent) of the structure to
increase linearly with radial distance. A CME of initially spherical
cross-section will thus tend to be flattened into a pancake-like shell, a
process counteracted only by the CME's own pressure-driven expansion and its
magnetic tension force, which acts towards a reduction of field line curvature
and thus tends to work against excessive flattening. \\
\citet{Jacobs-etal-2005} compared several popular CME propagation models and
found the resulting dynamics to be strongly dependent on both the included
physics (most notably the heating mechanism) and the background wind, among
other things confirming the intuitive notion that higher speeds are found in
fast, dilute winds. \citet{Chane-etal-2006} investigated the influence of the
CME's initial polarity with respect to the background field and found that
'normal' prominences result in fast, compact, approximately circular CMEs
which get deflected equatorwards, while CMEs which develop from 'inverse'
prominences are slower, pancake-shaped, and tend to deflect polewards, thus
confirming a previous theoretical prediction by \citet{Zhang-Low-2004}.

\subsection{The situation for 'realistic' models}

A functional prediction tool as envisaged by the space weather forecasting
community would take the observed distribution of photospheric magnetic field
at the instant of CME outbreak, infer from it the global initial state for all
MHD variable fields, then run a detailed CME simulation from which the
expected physical conditions at a given location (typically at Earth orbit)
can be extracted. Therefore, realistic models will typically try to reproduce
{\it in~situ} observations by spacecraft such as {\it WIND}, {\it Ulysses}, or
the {\it Advanced Composition Explorer} (ACE)
\citep[see, \eg,][]{Chane-etal-2008}. For regions closer to the Sun, another
possibility to assess the predictive capabilities of the model is the creation
of artificial white-light coronagraph images, which can be contrasted with
actual images of the respective event, as was done by, \eg,
\citet{Lugaz-etal-2007}. A first quantitative comparison of this type was
presented by \citet{Manchester-etal-2008}, who not only compared the absolute
projected brightness distribution but also the derived mass and velocity of
the CME in question. \\
While visual inspection of these comparisons often indicates impressive
agreement between simulation and observation, it must however be noted that
such comparisons tend to be biased towards good agreement, since publications
will typically show only a few selected results from a limited parameter
range, in spite of the fact that models are quite sensitive to the chosen
parameters \citep[\eg][]{Schrijver-etal-2008}. (For instance,
\citet{Chane-etal-2008} perform 24 different runs but explicitly state that
they only show their {\em best} fit to the ACE data of the target event.)
This makes it difficult to assess the predictive capabilities of a given model
for 'real world' forecasting applications, in which \mbox{\it a posteriori}
selection of parameter sets is not a valid option.

\section{Conclusions}
\label{s:conclusions}

CMEs represent a very diverse and important class of heliospheric transients.
Besides a purely scientific desire to understand their true nature, the urgent
need for accurate space weather forecasts creates a strong commercially-driven
incentive to intensify the study of these phenomena. \\
Since purely kinematic models are inadequate to capture their highly
non-linear evolution, 3-D MHD simulations are indispensable to model the
different stages of a CME's life cycle, with the long-term goal of providing
reliable forecasts. Associated difficulties mainly concern the need for high
spatial resolution --- although the ever-increasing performance of computing
power has helped much to mitigate this problem ---, the ambiguity induced by
the need to complement observational input by reasonable assumptions about the
remaining quantities not obtainable from direct observation, and the fact that
to this day, no universally accepted model for the initiation phase exists. \\
Generally speaking, the model developing community draws huge benefits from
high-quality spacecraft data to: 1. constrain the initial and boundary
conditions and 2. to allow for an {\it a posteriori} validation of the
obtained simulation results. \\
Since these results and predictions crucially depend on both the adopted
initial parameters and the physical effects included in the model, more
(parameter) studies are needed to quantify their respective influences.
Furthermore, an unbiased comparison between different space-weather prediction
tools is currently difficult because each group of authors tends to choose a
different event, relies on different input data from the event in question,
and sometimes also picks different methods and criteria to validate their
results. It thus seems that the community could benefit enormously from some
form of standardized space weather forecasting benchmark, in analogy to 
community-wide benchmark test suites known from and used for the simulation of
the heliosphere \citep{Mueller-etal-2008} or for convection in the Earth's
interior \citep{King-etal-2010}.

%
%
\begin{acks}
  Financial support through Research Unit 1048 (projects \mbox{FI~706/8-1}
  and \mbox{FI~706/8-2}), funded by the Deutsche Forschungsgemeinschaft
  (DFG), is gratefully acknowledged. This work also benefitted from the EU
  RNT {\em Solaire} (MTRN-CT-2006-035484). Furthermore, the author thanks
  Horst Fichtner and the anonymous referee for their constructive comments.
\end{acks}

%
%
\bibliographystyle{spr-mp-sola-cnd}
\bibliography{4pimodels}
%
%
%
%

\end{article} 
\end{document}